\documentclass[twocolumn]{aastex63}
\usepackage{graphics,epsf}
\usepackage{amsmath}                % American Mathematical Society package
\usepackage{amsfonts}               % American Mathematical Society fonts
\usepackage{amssymb}                % American Mathematical Society symbol
\usepackage{epsfig}                 % EPS figures
\usepackage{appendix}
\usepackage{graphicx}
\usepackage{float}
\usepackage{color}
\usepackage{multirow}
\usepackage{colortbl}
\usepackage[para,online,flushleft]{threeparttable}

\hypersetup{citecolor=blue, % color for \cite{...} links
            linkcolor=red, % color for \ref{...} links
            menucolor=blue, % color for Acrobat menu buttons
            urlcolor=blue}  % color for \url{...} links

\newcommand{\cm}{{~\rm cm}}
\newcommand{\m}{{~\rm m}}
\newcommand{\km}{{~\rm km}}
\newcommand{\s}{{~\rm s}}
\newcommand{\h}{{~\rm h}}

\newcommand{\g}{{~\rm g}}
\newcommand{\G}{{~\rm G}}
\newcommand{\K}{{~\rm K}}
\newcommand{\erg}{{~\rm erg}}

\begin{document}

\title{Implications of post-kick jets in core collapse supernovae} 

%% \correspondingauthor{Noam Soker}
%% \email{soker@physics.technion.ac.il}

%% \author{Efrat Sabach}
%%% \affiliation{Department of Physics, Technion, Haifa, 3200003, Israel}

\author[0000-0003-0375-8987]{Noam Soker}
%\affil{Departmeמt of Physics, Technion, Haifa 3200003, Israel}
\affiliation{Department of Physics, Technion, Haifa, 3200003, Israel; soker@physics.technion.ac.il}

%% \author{Rbert T. Fisher}
%% \affiliation{Department of Physics, University of Massachusetts  Dartmouth, 285 Old Westport Road, North Dartmouth, MA 02740, USA; rfisher1@umassd.edu}

\begin{abstract}
I examine the assumption that the jets that shape the axisymmetrical morphological features of core collapse supernova (CCSN) remnants are post-kick jets, i.e., the neutron star (NS) launches these jets after the explosion and after it acquired its natal kick velocity. I find that this assumption implies that the pre-collapse cores of CCSN progenitors have sufficient angular momentum fluctuations to support jittering jets that explode the star. From the finding that the shaping-jets neither tend to be aligned with the kick velocity nor to be perpendicular to it I argue that the assumption that the shaping-jets are post-kick jets has the following implications. (1) The NS accretes mass at a radius of $r_{\rm acc} \approx 5000 \km$ from the center of the explosion at $\approx 10 \s$ after explosion.  (2) The required angular momentum fluctuations of the accreted gas to explain the medium values of jets-kick angles are also sufficient to support an intermittent pre-kick accretion disk, just before and during the explosion. Such an intermittent accretion disk is likely to launch jets that explode the star in the frame of the jittering jets explosion mechanism. This suggests that most likely the shaping-jets are the last jets in the jittering jets explosion mechanism rather than post-kick jets. (3) The jittering jets explosion mechanism expects that black holes have small natal kick velocities.  
\end{abstract}


% ==========================================================
\section{Introduction}
\label{sec:intro}
% ==========================================================

A large fraction of neutron stars (NSs) acquire a natal kick velocity during their formation process in a core-collapse supernova (CCSN) explosion, with typical velocities of $v_{\rm NS} \simeq 200 - 500\km \s^{-1}$ and up to $\simeq 1000\km \s^{-1}$
(e.g., \citealt{Cordesetal1993,LyneLorimer1994, Chatterjeeetal2005, Hobbsetal2005, Kapiletal2022}).
The natal kick velocity most likely results from non-spherical explosion geometry (e.g., \citealt{Laietal2006, Wongwathanaratetal2013, Janka2017, HollandAshfordetal2017, Katsudaetal2018}). Some other proposed mechanisms to account for natal kick velocity suffer from difficulties (e.g., \citealt{Lai2003, Wongwathanaratetal2010, Nordhausetal2010, Nordhausetal2012, Katsudaetal2018}). Combination of more than one asymmetrical mechanism to impart natal kick velocity is possible, e.g., \cite{Wangetal2006} who discuss the combination of rapid rotation and magnetic fields to launch jets that might impart a kick velocity. In the latter scenario the jets and the kick velocity tend to be aligned, in contradiction to the finding that I summarize in section \ref{sec:kick}. 

Many studies concentrate on the relation between the NS spin direction and kick velocity direction (e.g., \citealt{SpruitPhinney1998, Laietal2001, Dodsonetal2003, Johnstonetal2005, FryerKusenko2006, Johnstonetal2006, NgRomani2006, NgRomani2007, Wangetal2007, Kaplanetal2008, BrayEldridge2016, MandelIgoshev2022}).
Some studies expect that in the delayed neutrino explosion mechanism  the kick velocity and the NS spin axis be at small angles with respect to each other, i.e., be aligned (e.g., \citealt{Wongwathanaratetal2013, Jankaetal2022Spin}). In a recent study \cite{Jankaetal2022Spin} propose that tangential vortex flows of the gas that the NS accretes after it acquires its kick velocity can explain spin-kick alignment. 

Here I will follow other studies (e.g., \citealt{BearSoker2018kick, Soker2022SNR0540}) that explore the relation between the direction of the jets' axis, i.e., the axis along the two opposite jets, and the kick velocity direction. These studies assume that two late opposite jets shaped some CCSN remnants (CCSNRs) that possess axisymmetrical morphologies, and take the jets' axis to be the line connecting the two opposite ears in some CCSNRs or the axis along the faint elongated inner part of the SNR. As well, I will follow \cite{Soker202187Akick} and \cite{Jankaetal2022Spin} in considering post-explosion accretion of mass with large amount of angular momentum onto the newly born NS. However, instead of considering the final spin of the NS as a result of this post-explosion accretion \citep{Jankaetal2022Spin}, I consider the role of this accretion in determining the power and direction of the final jet-launching episode.  Beside that, the present study has a lot in common with the assumptions and studied processes in the study by \cite{Jankaetal2022Spin}. 

Both the neutrino delayed and the jittering jets explosion mechanisms allow for post-explosion jets. For example, \cite{Orlandoetal2021} consider the delayed neutrino explosion mechanism for Cassiopeia A. They suggest that because instabilities in the delayed neutrino explosion mechanism (see, e.g., \citealt{Wongwathanarat2017, Utrobinetal2019, Jerkstrandetal2020, Orlandoetal2020}) by themselves cannot explain all properties of this CCSNR, post-explosion jets shaped the features of jets in the Cassiopeia A CCSNR. The jittering jets explosion mechanism where mainly jets power the explosion (e.g., \citealt{PapishSoker2011, GilkisSoker2016, ShishkinSoker2022}), allows also for late accretion and then launching of late jets, e.g., as I suggested for SN 1987A \citep{Soker202187Akick}. 

In section \ref{sec:AngularMomentum} I explore some possible implications of the medium ($\approx 50^\circ$) angles between the jets' axis and the kick velocity. Section \ref{sec:AngularMomentum} contains the main new results of this study. I first present this medium angle distribution in section \ref{sec:kick} and discuss in section \ref{sec:TugBoat} the tug-boat acceleration mechanism that I take to accelerate  the NS to its natal kick velocity (e.g., \citealt{Schecketal2004, Schecketal2006, Nordhausetal2010, Wongwathanaratetal2010, Wongwathanaratetal2013, Janka2017}; I note that there are other proposed mechanisms, e.g., \citealt{Yaoetal2021} and \citealt{Xuetal2022}). I summarize the main new results in section \ref{sec:Summary}. 

%==========================================================
\section{Kick velocity versus jets direction}
\label{sec:kick}
% ==========================================================

I deal here with the projected angle $\alpha$ on the plane of the sky between the NS velocity and the axis of the main pair of opposite jets. \cite{BearSoker2018kick} estimate this angle for 12 SNRs and  \cite{Soker2022SNR0540} for one SNR. They determine the jets’ axis by assuming that the jets shape axially symmetric CCSNRs and that the symmetry axis of a morphology is the jets' axis. Such features might be two opposite protrusions from the main CCSNR shell termed ears or a faint elongated structure through the center of the CCSNR. These papers present the cumulative distribution function ${\rm W}_{\alpha}({\rm obs})$ of the projected angles $\alpha$ for these CCSNRs.
 
\begin{figure}[!t]
%\centering
%\vskip -2.00cm
%\hskip -2.00cm
\includegraphics[trim=3.6cm 8.2cm 0.0cm 8.0cm ,clip, scale=0.63]{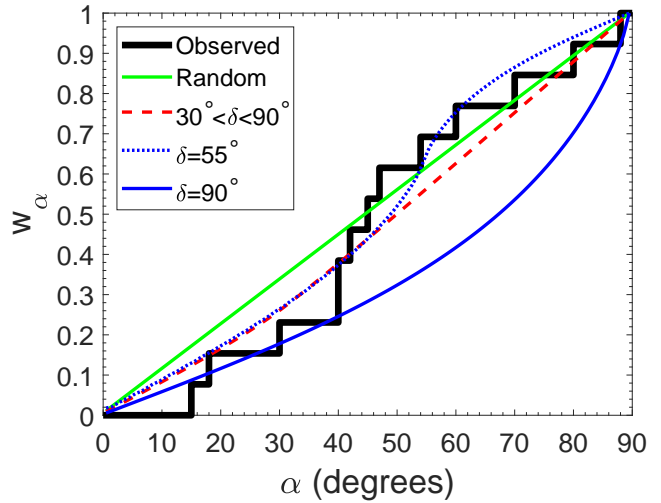}
%\vskip -3.9cm
\caption{The observed cumulative distribution function ${\rm W}_{\alpha}({\rm obs})$ of the projected angles $\alpha$ between the jets'-axis and the NS natal kick velocity for 13 SNRs (black step function; 12 SNRs from \citealt{BearSoker2018kick} and one from \citealt{Soker2022SNR0540}). Other lines represent theoretical expectations. The straight green line represents the cumulative distribution function ${\rm W}_{\alpha}({\rm rand})$ for a random three-dimensional angle between the kick and jet directions, the convex blue line represents the cumulative distribution function ${\rm W}_{\alpha}(90)$ when in all cases the three-dimensional NS kick velocity is perpendicular to the jets' axis, i.e., $\delta=90^\circ$, the dashed-red line is the cumulative distribution function ${\rm W}_{\alpha}(30-90)$ when the probability for the angle $\delta$ is by equation (\ref{eq:deltaAngle}) with $\delta_c=30^\circ$, and the dotted-blue line that curves around the green line is the cumulative distribution function $W_\alpha(55)$ for a fixed angle of $\delta=55^\circ$ between the kick and jet directions. 
}
\label{fig:distribution}
\end{figure}

In Fig. \ref{fig:distribution} I present by a thick-black step function the observed cumulative distribution function ${\rm W}_{\alpha}({\rm obs})$ for the 13 SNRs from \cite{Soker2022SNR0540}. 
 The small number of objects, only 13, implies that the observed cumulative distribution function is highly uncertain. This should be kept in mind in analysing the implications of Fig. \ref{fig:distribution}.  
Other lines in Fig. \ref{fig:distribution} represent different theoretical expectations. The straight green line represents the cumulative distribution function ${\rm W}_{\alpha}({\rm rand})$ when the three-dimensional angle $\delta$ between the kick velocity and the jets' axis is random. The convex blue line represents the cumulative distribution function ${\rm W}_{\alpha}(90)$ when in all cases $\delta=90^\circ$ (but random in its $360^\circ$ direction around this angle).  
%FFFFFFFFFFFFFFFFFFFFFFFFFFFFFFFFFFFFFFFFFFFFFFFFFFFFFFFFFFFFFFFFFF
% FFFFFFFFFFFFFFFFFFFFFFFFFFFFFFFFFFFFFFFFFFFFFFFFFFFFFFFFFFFFFFFFFF

  At large angles of $\alpha \ga 45^\circ$ the observed cumulative distribution function is similar to the random distribution ${\rm W}_{\alpha}({\rm rand})$, while at small angles of $\alpha \le 40 ^\circ$ it is more similar to the perpendicular distribution ${\rm W}_{\alpha}(90)$ \citep{BearSoker2018kick, Soker2022SNR0540}. This actually implies that neither the random distribution function W(rand) nor the perpendicular distribution function W(90) fit the observed cumulative distribution function. 

Here I go further and consider the cumulative distribution function for a three-dimensional kick-jet angle distribution that is random for $\delta_c < \delta \le 90^\circ$. The probability for an angle $\delta$ to be in the range $\delta$ to $\delta + d \delta$ is therefore
\begin{equation}
P(\delta) d \delta = \begin{cases}
        0  & \delta \leq \delta_c
        \\
        \frac{\sin \delta}{\cos \delta_c} d \delta 
        & \delta_c < \delta \le 90^\circ  . 
        \end{cases}
\label{eq:deltaAngle}
\end{equation}
    \newline
I draw the cumulative distribution function $W_\alpha(30-90)$ for $\delta_c=30^\circ$ by a dashed-red line. I take $\delta_c \simeq 30^\circ$ that best fits the observed distribution, i.e., it gives the minimum value of the maximum distance between $W_\alpha(\delta_c-90)$ and $W_\alpha({\rm obs})$. I also present, by the dotted-blue line, the cumulative distribution function for a fixed three-dimensional angle of $\delta=55^\circ$ between the kick and jets directions $W_\alpha(55)$ (it is random in its $360^\circ$ direction around that angle).  I use $W_\alpha(55)$ because it fits better the observed cumulative distribution function than a fixed angle of $\delta \la 50^\circ$ or $\delta \ga 60^\circ$. I do not attribute a particular meaning to the value of $\delta=55^\circ$, as there are uncertainties in the observed cumulative distribution function. This value simply implies that the angle $\delta$ avoids small values, like the function $W_\alpha(30-90)$ also implies.  The functions $W_\alpha(30-90)$ and $W_\alpha(55)$ better fit the observed cumulative distribution function than  ${\rm W}_{\alpha}({\rm rand})$ and $W_\alpha(90)$ do.  

The point of this section is that the typical component of the kick velocity along the jets' axis ${v_{\rm NS} {\cos \delta}}$ is of similar value to the typical kick velocity component perpendicular to the jets' axis ${v_{\rm NS} {\sin \delta}}$, namely, 
\begin{equation}
{\tan \delta} \approx 1.  
\label{eq:angles} 
\end{equation}
 The range of delta can be $30 \la \delta \la 60^\circ$, implying $0.6 \la \tan \delta \la 1.7$. 
I will discuss possible implications of the approximate equality (\ref{eq:angles}) in section \ref{sec:AngularMomentum}. I first discuss the tug-boat mechanism to accelerate the NS to its kick velocity. 

%====================================================
\section{Using the gravitational tug-boat mechanism}
\label{sec:TugBoat}
%====================================================

\cite{Wongwathanaratetal2013} analytically estimate the natal kick velocity in the tug-boat mechanism, i.e., where an ejecta clump (or several close clumps)  gravitationally pulls the NS and accelerates it to its natal kick velocity. They take the extra ejecta mass on one side to be $\Delta m \simeq 10^{-3} M_\odot$ and an equal mass that is missing on the other side. They take the ejecta shell to expand with a constant velocity of $v_{\rm s}=5000 \km \s^{-1}$, and they take the acceleration to start when the shell is at a radius of $r_{\rm i}=100 \km$ from the NS. 
The assumption that the shell expands at a constant velocity of $v_{\rm s}=5000 \km \s^{-1}$ is a very strong assumption because the expected expansion velocity of the inner ejecta is much smaller. We must consider that the NS slows down the shell, in particular the part that pulls the NS as the NS is closer to that part of the shell.  

\cite{Wongwathanaratetal2013} discuss also the case where instead of mass asymmetry the gravitational tug-boat mechanism can operate with a slower shell segment. The slower shell segment has more interaction time with the NS and closer distances, and hence it pulls the NS.  

The gravitational tug-boat acceleration is a long-lasting (many seconds) process where internal energy of the ejecta is converted into kinetic energy of the ejecta that maintains the shell expansion \citep{Wongwathanaratetal2013}. This might be compatible with the several seconds explosion timescale that some new studies of the delayed neutrino mechanism find (e.g., \citealt{Imashevaetal2022}). 
It is also compatible with the jittering jets explosion mechanism. 

However, at later times the acceleration of the shell declines, and the gravitational pull of the NS decelerates the slow shell segment to the degree that the NS might `collide' with this mass, or, more accurately, the NS accretes this mass. I now try to estimate plausible parameters for this accretion. 

First I take the asymmetrical mass distribution to be somewhat larger than what \cite{Wongwathanaratetal2013} take, i.e., $\Delta m \simeq 0.001-0.01 M_\odot$, and so I scale the post-kick accreted mass with $M_{\rm acc,pk} \simeq 0.003 M_\odot$ in what follows. The power of the two jets together is 
\begin{eqnarray}
\begin{aligned} 
\dot E_{\rm 2j} & = \eta M_{\rm acc,pk} \frac{G M_{\rm NS}}{R_{\rm NS}}
= 9.2 \times 10^{49} 
\\ & \times \left( \frac{\eta}{0.1} \right)
\left( \frac{M_{\rm acc,pk}}{0.003M_\odot} \right) \erg  ,
\label{eq:E2jA}
\end{aligned}
\end{eqnarray}
where $\eta \ll 1$ is an efficiency parameter as most accretion power is carried by neutrinos, and I take throughout this study $M_{\rm NS}=1.4M_\odot$ and $R_{\rm NS}=12 \km$ for the NS mass and radius, respectively. Note that the NS is already (almost) relaxed, as the post-kick accretion takes place many seconds ($t>10 \s$) after explosion.  For a typical total energy budget of a CCSN the neutrinos carry more than 99 per cent of the energy that the collapsing core releases. However, the flow structure here is not of a spherical collapse but rather of an accretion disk that launches jets. For that, the fraction of energy that the jets carry is typically much larger than one per cent, as the typical fraction in other astrophysical  objects that launches jets via an accretion disk.  In \cite{Soker202187Akick} I suggested that the NS remnant of SN 1987A accreted a mass of $\approx 2 \times 10^{-4} - 0.002 M_\odot$ for an efficiency parameter $\eta=0.1-0.01$, respectively. Here I consider much more powerful jets because I consider jets that shape out to the outskirts of the ejecta of the CCSNR. In SN 1987A the assumed post-kick jets shaped only a small region within the ejecta. 

The energy of the jets according to the scaling of equation (\ref{eq:E2jA}) is the typical energy of jets that \cite{GrichenerSoker2017} argue to inflate ears in CCSNRs.

I take a second approach to estimate the accreted mass, which although crude, might give an indication to plausible values of the accreted mass. The main drawback of this approach is that, as I discuss in section \ref{sec:AngularMomentum}, I expect accretion to take place before the ejecta reaches homologous expansion. Nonetheless, I find this approach to be relevant to present it here in short. Consider then that the NS acquires a kick velocity of $v_{\rm NS}$ and that it catches up with the ejecta that already acquired its homologous expansion. I take a spherical ejecta density profile from \cite{SuzukiMaeda2019} with $\delta=1$ and $m=10$ (their equation 1-6 based on \citealt{ChevalierSoker1989}) 
\begin{equation}
\rho (r, t) = \begin{cases}
        \rho_0 \left( \frac{r}{t v_{\rm br}} \right)^{-1} 
        & r\leq t v_{\rm br}
        \\
        \rho_0 \left( \frac{r}{t v_{\rm br}} \right)^{-10} 
        & r>t v_{\rm br}, 
        \end{cases}
\label{eq:density_profile}
\end{equation}
    \newline
where $M_{\rm ej}$ is the ejecta mass, $E_{\rm SN}$ is its kinetic energy,  
\begin{eqnarray}
\begin{aligned} 
& v_{\rm br} = \left( \frac{20}{7} \right)^{1/2} \left( \frac {E_{\rm SN}}{M_{\rm ej}} \right)^{1/2} 
=3.79 \times 10^3 % 3923.1 
\\& 
\times \left( \frac {E_{\rm SN}}{10^{51} \erg} \right)^{1/2}
\left( \frac {M_{\rm ej}}{10 M_\odot} \right)^{-1/2}
\km \s^{-1} ,
\end{aligned}
\label{eq:vbr}
\end{eqnarray}
and
\begin{equation}
\rho_0 = \frac {7 M_{\rm ej}}{18 \pi v^3_{\rm br} t^3} .
\label{eq:rho0}
\end{equation}
The ejecta mass from the center and up to the ejecta velocity with which the NS interacts, $v_{\rm ej}=v_{\rm NS}$, is 
\begin{eqnarray}
\begin{aligned}
M_{\rm ej} &  (v_{\rm NS}) = \int^{r(v_{\rm NS})}_0 \rho_0 \left[ \frac{r}{r(v_{\rm br})} \right]^{-1} 4 \pi r^2 dr 
\\ &
= \frac{7}{9} \left( \frac{v_{\rm NS}} {v_{\rm br}} \right)^2 M_{\rm ej}=   0.12 \left( \frac {M_{\rm ej}}{10 M_\odot} \right)
\\ & \times 
\left( \frac {v_{\rm NS}}{500\km \s^{-1}} \right)^{2}
\left( \frac {v_{\rm br}}{4000 \km \s^{-1}} \right)^{-2}
M_\odot. 
\end{aligned}
\label{eq:M(lobe)}
\end{eqnarray}
The NS accretes only ejecta mass from the inner region that is on the side of its kick velocity and close to it. To obey the scaling of equation (\ref{eq:E2jA}) the NS should accrete $\simeq 2.5\%$ of the mass of the shell it catches up with according to the scaling of equation (\ref{eq:M(lobe)}), i.e., $M_{\rm acc,pk} \simeq 0.025 M_{\rm ej}  (v_{\rm NS})$.  Note that the gravitational radius of influence from which the NS accretes mass is much larger than the NS size, being $\simeq 10^4 \km$ (e.g., \citealt{Jankaetal2022Spin}) for the parameters of this study.  For a faster kick velocity and/or a larger ejecta mass the required fraction of accreted ejecta mass is smaller. The purpose of equation (\ref{eq:M(lobe)}) is only to show that a post-kick accreted mass of $M_{\rm acc,pk} \simeq 0.003 M_\odot$ is reasonable.

The main point of this section is that post-kick accretion can lead to jets that inflate the ears that are observed in some CCSNRs. This does not mean that in all cases the jets that shape ears are late (several to tens of seconds after explosion) post-kick jets. In some cases, or even all cases, these might be just the last jets in the jittering jets explosion mechanism (e.g., \citealt{GrichenerSoker2017, Bearetal2017}). In this study I explore the implications of the assumption that post-kick jets shape CCSNRs, e.g., inflating the ears. 

%====================================================
\section{Angular momentum considerations}
\label{sec:AngularMomentum}
%====================================================

In this section I consider the accretion of material with angular momentum by the NS after it acquired its kick velocity (post-kick accretion). I consider this material to form an accretion disk (or an accretion belt, i.e., a thick disk supported not only by centrifugal forces; \citealt{SchreierSoker2016, GarainKim2023}) that launches two opposite jets along the angular momentum axis of the accretion disk.  
%=========================
\subsection{General considerations}
 \label{subsec:GeneralConsiderations}
%=========================
As I mentioned in section \ref{sec:intro}, \cite{Jankaetal2022Spin} consider post-kick accretion onto the NS to explain the alignment of the NS spin with its kick velocity. In their picture the source of angular momentum is the presence of vortexes in the accreted gas.  \cite{Jankaetal2022Spin} scale their equations with a kick velocity of $500 \km \s^{-1}$, with a radius (relative to the center) where accretion takes place of $3 \times 10^4 \km$, and an accreted mass of $< 0.0025 M_\odot$. The accretion therefore occurs within a minute to a few minutes after explosion. To account for the spin-kick alignment \cite{Jankaetal2022Spin} consider vortexes in the accreted gas with radial angular momentum, namely, along the kick velocity in the region where accretion takes place. They did not consider the angular momentum component due to non-axisymmetrical accretion geometry that I study in section \ref{subsec:MediumAngles}.  

Contrary to the claim of spin-kick alignment, I here consider jet-kick misalignment (Fig. \ref{fig:distribution}). Namely, the jets-axis tends to avoid small angles  to the kick velocity, or, as I conclude in section \ref{sec:kick}, on average the component of the kick velocity along the jets' axis has a similar value to the kick velocity component perpendicular to the jets' axis (equation \ref{eq:angles}). Here I attribute jet-kick misalignment to stochastic velocities of the accreted mass, in the spirit of the jittering jets explosion mechanism.  

Earlier papers \citep{BearSoker2018kick, Soker2022SNR0540} attributed the medium kick-jet angles ($\delta \approx 50^\circ$) to the tug-boat acceleration mechanism that operates in the jittering jets explosion mechanism. 
I considered two possibilities of the relation between the dense ejecta clumps that accelerate the NS and the exploding jets. In the first explanation the jets prevent the formation of dense clumps along their propagation directions and therefore no NS acceleration takes place in those directions. In the second explanation several dense clumps are falling from about the same direction. Some clumps feed the NS and some escape and accelerate the NS. Namely, in addition to accelerating the NS by some dense clumps, other dense clumps from the same direction also supply the gas to the accretion disk that launches the last pair of jets. Therefore, the last pair of jets tend to be misaligned with the direction of the clumps, which is the acceleration direction.  

In this study I differ from these explanations and examine the implications of the assumption that the NS launches the last pair of jets after the explosion already took place. The explosion could be driven by the delayed neutrino mechanism or by the jittering jets explosion mechanism. Namely, the post-kick jets are not related directly to the explosion process. \cite{Orlandoetal2021}, for example, suggest that in Cassiopeia A the NS launched the jets after the explosion, that they take to be the delayed neutrino mechanism. I discuss the implications of this assumption in relation to the jet-kick misalignment, i.e., the medium kick-jet angles as I give in equation (\ref{eq:angles}). 
 
%=========================
\subsection{Medium jet-kick angles by post-kick accretion}
 \label{subsec:MediumAngles}
%=========================

I first assume that the accretion of mass by the NS from a dense clump forms a long-lasting accretion disk that launches pair of jets with a fixed axis. I will return to this assumption at the end of this subsection. 

I consider that the NS moves through a dense clump and accretes from it. I take this clump to be inhomogeneous and decompose it to $N_{\rm b}$ blobs with a typical mass per blob of $m_{\rm b}$, and the typical impact parameter of a blob (distance of the blob from the line along the NS motion) is $D_{\rm b}$. I consider a small number $1< N_{\rm b} <10$ of blobs. The properties of blobs are determined by the pre-collapse convective motion (vortexes), as I assume that these blobs are descendent of convective cells. 
 I estimate the typical size of a convective cell to be the mixing length in the pre-collapse core. The ratio of the mixing length to the radius in a pre-collapse core is $\simeq 0.4$ (e.g., \citealt{ShishkinSoker2021}). 
In a volume of $(r)^3$ from which the NS accretes mass, therefore, there are $\simeq 10-20$ blobs. The NS accretes from a smaller volume even, and so I scale with a number of $N_{\rm b} =5$ blobs.   

The typical specific angular momentum of a blob is therefore $j_{\rm b}=D_{\rm b} v_{\rm NS}$, assuming that the blobs are at rest or have very small velocities relative to the center of mass of the progenitor. I also assume that the blobs that the NS accretes are distributed at random with respect to the direction of motion. The specific angular momentum of the total accreted mass is given by a random addition of the angular momenta of the blobs. The value is, under the assumption that all blobs have the same mass in this toy-model construction,  
\begin{eqnarray}
\begin{aligned} 
j_{\rm acc,kick} & \approx \frac{1}{\sqrt{N_{\rm b}}} j_{\rm b} \simeq 10^{16} \left( \frac{N_{\rm b}}{5} \right)^{-1/2} 
\\ & 
\times \left( \frac{D_{\rm b}}{5000 \km} \right)  
\left( \frac{v_{\rm NS}}{500 \km \s^{-1}} \right)  \cm^2 \s^{-1}. 
\label{eq:jacckick}
\end{aligned}
\end{eqnarray}
For a typical average natal kick velocity I take $v_{\rm NS} = 500 \km \s^{-1}$ (e.g., \citealt{Hobbsetal2005, Kapiletal2022}), and I scale with the impact parameter of $5000 \km$ which is about the radii of the inner core layers that are expelled in the explosion.  
The specific angular momentum of an object on a circular orbit on the surface of a NS is $j_{\rm circ} \simeq 2 \times 10^{16} \cm^2 \s^{-1}$. Although the average specific angular momentum as given by the scaling of equation (\ref{eq:jacckick}) is too low to form a thin disk, it is large enough to form a thick accretion belt that also might launch jets (e.g., \citealt{SchreierSoker2016}).   

The explosion by itself imparts mainly radial velocity to the blobs that compose the dense clump and to the NS. The relative blobs-NS velocity is radial, and therefore the angular momentum of each blob is perpendicular to the kick velocity, which is also radial. The sum of these angular momenta, hence the jets' axis, will also be perpendicular to the kick direction, i.e., $\delta=90^\circ$, had it been only due to the kick velocity. However, there is another source of angular momentum of the blobs, the random motion of the gas due to the pre-collapse convective motion in the core, or vortexes as \cite{Jankaetal2022Spin} propose. 

I mark by $j_{\rm acc,conv}$ the total specific angular momentum of the accreted blobs as a result of pre-explosion convective motion. Since the jets are launched along the direction of the combined angular momenta $\vec{j}_{\rm acc,kick}+\vec{j}_{\rm acc,conv}$  and $\vec{j}_{\rm acc,kick}$ is perpendicular to the kick velocity,  the minimum value of the jet-kick angle $\delta$ is obtained when $\vec{j}_{\rm acc,conv}$ is along the kick direction. 
Using the estimate from  equation (\ref{eq:angles}) I conclude that the typical value of the pre-collapse stochastic specific angular momentum of the post-kick accreted mass is 
\begin{equation}
j_{\rm acc,conv} \ga (\tan \delta )^{-1} j_{\rm acc,kick}.  
\label{eq:jacccon} 
\end{equation}

The impact parameter of each blob when we consider its random velocity due to convection is different from the impact parameter $D_{\rm b}$ due to the kick velocity, but not by much. The average value will be about the same $D_{\rm b,conc} \simeq D_{\rm b}$. The number of accreted blobs is the same of course. We can therefore use an expression similar to equation (\ref{eq:jacckick}) but for the random convective velocity of the blobs rather than the kick velocity, i.e., replacing $v_{\rm NS}$ by $v_{\rm conv}$ in equation (\ref{eq:jacckick}). Equation (\ref{eq:jacccon}) implies then that
\begin{equation}
v_{\rm conv} \ga v_{\rm NS} \approx 500 \km \s^{-1}.  
\label{eq:velocities} 
\end{equation}
This value is compatible with the convection velocity in the silicon and oxygen burning shells of pre-collapse cores, e.g., \cite{ShishkinSoker2021} who argue that this is sufficient to account for jittering jets that explode the star. \cite{ShishkinSoker2021} find that the mixing length theory yields pre-collapse convective velocities of $\simeq 200 \km \s^{-1}$ (like in the oxygen burning shell at $r \simeq 2500 \km$), but note that three-dimensional simulations, e.g., by \cite{FieldsCouch2021}, give velocity amplitudes that are three to four times as large. 

If the accretion by the NS takes place at much larger radii $r \gg 10^4 \km$ then the impact parameter in equation (\ref{eq:jacckick}) $D_{\rm b}$ would be larger and so is the specific angular momentum due to the kick velocity $j_{\rm acc,kick}$. However, if the oxygen burning convective layer moves out to large distances and each convective cell conserves its angular momentum the tangential random velocity of the cells would be reduced to the degree that inequality (\ref{eq:velocities}) would be violated. As well, outer convective regions in the core have lower convective velocities that what equation (\ref{eq:velocities}) requires (lower than in the oxygen burning shell).  The implication is that to obey equation (\ref{eq:angles}) the accretion to launch the jets under the present assumptions should occur at a radius of $r_{\rm acc} \approx 5000 \km$ (possibly within the range of $r_{\rm acc} \simeq 2000-2\times 10^4 \km$). This is the first conclusion of this section.   

The combined specific angular momenta of the convective cells decreases with increasing number of convective cells. The post-kick accretion is from a limited volume of a given shell, i.e., the zone along the direction of the NS kick, and therefore involves a small number of convective cells. On the other hand, the accretion process onto the NS before it acquires its kick velocity, namely during the explosion process, involves the entire shell and hence involves more convective cells. This implies specific angular momentum that is lower than the value as given by equation (\ref{eq:jacccon}). However, instabilities behind the stalled shock at $r \la 100 \km$ during the explosion process increase the stochastic angular momentum amplitudes (\citealt{Soker2019SASI, Soker2019JitSim}). Instabilities include the spiral standing accretion shock instability (SASI; for the spiral-SASI see, e.g., \citealt{Andresenetal2019, Walketal2020, Nagakuraetal2021, Shibagakietal2021}). This brings me to the second conclusion of this section. 

The second conclusion, under the assumption that the jets that shape CCSNRs are post-kick jets, is that the jets-kick misalignment (equation \ref{eq:angles}) implies that the convective stochastic motion in the pre-collapse core is large enough to form an intermittent accretion belt or disk that launches jets with varying direction and power. These jets explode the star according to the jittering jets explosion mechanism. 

In this section I assumed that the accretion of gas from the inhomogeneous clump (that I decomposed to blobs) leads to a fixed-axis jets, i.e., very small jittering. However, the dynamical time of the accretion disk around the NS is very short, much less than a second. The accretion process lasts for ten seconds or more. Therefore, it is very likely that the random motion of the blobs and their random directions with respect to the NS will lead to jittering jets. Therefore, it is not clear at all that post-kick accretion can lead to a fixed-axis jets, or even jets with small jittering. Nonetheless, above I examined the implications of the assumption that post-kick accretion can lead to the launching of jets with small jittering. 

%===========================================
\section{Summary}
\label{sec:Summary}
%===========================================

Many CCSNRs possess morphological features that suggest shaping be jets. Some studies consider the shaping-jets to be the last pair of jets in the jittering jets explosion mechanism (e.g., \citealt{GrichenerSoker2017, Bearetal2017}). On the other hand, \cite{Orlandoetal2021} who take the neutrino delayed explosion mechanism to have exploded Cassiopeia A suggest that the shaping of the ears in that CCSNR was by jets that the NS launched after explosion. At that time the NS already acquired its natal kick velocity, i.e., these are post-kick jets.  In this paper I studied the implications of the assumption that the axisymmetrical morphological features in CCSNRs are due to post-kick jets.

I first reanalyzed the distribution of projected (on the plane of the sky) angles between the inferred jets' axis and the kick velocity (section \ref{sec:kick}), and examined two simple fittings to the observed cumulative distribution function of the projected angles (Fig. \ref{fig:distribution}). These fittings suggest that the typical (three-dimensional rather than projected) angles between the kick velocity and the shaping-jets are scattered around $\delta \approx 50^\circ$, which I quantify by equation (\ref{eq:angles}). If the accreted angular momentum was only due to the NS kick velocity (equation \ref{eq:jacckick}) the angle between the jets and the kick velocity would have been $\delta = 90^\circ$. The medium values that I infer, $\delta \approx 50^\circ$, implies that there is another source of angular momentum. I here take this source of angular momentum to be the convective motion in the accreted gas. Namely, accretion from a region that was part of convective shell in the pre-collapse core. 

The accreted clump might be the same clump that accelerates the NS to its natal kick velocity by the tug-boat mechanism that I discussed in section \ref{sec:TugBoat}. In section \ref{sec:AngularMomentum} I examined the implications of the finding that $\tan \delta \approx 1$ (equation \ref{eq:angles}) under the assumption that post-kick jets shape CCSNRs and that the accretion leads to the launching of post-kick jets with a fixed axis (only small jittering). This fixed axis is required to explain an axisymmetrical structure of the CCSNRs that I consider here. It is not clear at all that the post-kick accretion can form fixed-axis post-kick jets. 

I note that the formation of post-kick jets from the oxygen burning layer might imply the presence of large fractions of silicon and sulfur in the jets. This might explain the high abundance of Si/S in the jets of Cassiopeia A (for observations, see, e.g., \citealt{FesenMilisavljevic2016, Grefenstetteetal2017}). 
   
I can summarize my main conclusions as follows, where the first two conclusions are from section \ref{subsec:MediumAngles}, and the third is new to this section. 
\begin{enumerate}
\item  If post-kick jets shape CCSNRs, then the post-kick accretion takes place at $r_{\rm acc} \approx 5000 \km$ (accretion might take place in the range of $r_{\rm acc} \simeq 2000-2\times 10^4 \km$).
\item The angular momentum fluctuations in the accretion region are sufficiently large to allow the formation of post-kick accretion belt or accretion disk. This implies that the convective velocity fluctuations in the pre-collapse core allow also for intermittent disk formation during the explosion itself. Namely, the assumption of post-kick jets and the finding that they are at medium angles to the kick direction suggest the launching of jets during the explosion process. This supports the jittering jets explosion mechanism. 
\item In cases of a bipolar explosion that are driven by fixed-axis jets (jittering is very small) there are no jets in and close to the equatorial plane (perpendicular to the jets' axis) during the explosion process. Therefore, large amounts of mass are accreted from the equatorial plane region and hence there will be no ejected clumps to gravitationally poll the newly born NS or black hole in the tug-boat mechanism. The expectation of the jittering jets explosion mechanism is that black holes and possibly massive NSs have very small natal kick velocities.  
\end{enumerate}

Overall, the questionable assumption that the post-kick accretion can form fixed-axis jets and implication 2 above suggest that most likely the shaping of axisymmetrical features in CCSNRs is by the last jets of the jittering jets explosion mechanism. 

%===========================================
\section*{Acknowledgments}
%===========================================
 I thank Thomas Janka and an anonymous referee for useful and clarifying comments.  
This research was supported by a grant from the Israel Science Foundation (769/20). 
% =======================

%%%%%%%%%%%%%%%%%%%%%%%%%%%
\section*{Data availability}
The data underlying this article will be shared on reasonable request to the corresponding author. 
%%%%%%%%%%%%%%%%%%%%%%%%%%%

\end{document}